\documentclass[amssymb,amsmath,footinbib,rmp,twocolumn]{revtex4}
\usepackage{natbib,graphicx,subfigure,subeqnarray,fancyhdr,amsmath,multirow,color, mathrsfs}
\begin{document}

\def\De{{\rm De}}

\title{Bacterial Hydrodynamics\footnote{to appear in: Annu.~Rev.~Fluid Mech.~{\bf 48} (2016)}}
\author{Eric Lauga\footnote{e.lauga@damtp.cam.ac.uk}}
\affiliation{Department of Applied Mathematics and Theoretical Physics, 
Centre for Mathematical Sciences,  University of Cambridge, 
Wilberforce Road, Cambridge, CB3 0WA, United Kingdom}
\date{\today}

\begin{abstract}
Bacteria  predate  plants and animals by billions of years. Today, they are the world's smallest cells yet they represent the bulk of the world's biomass, and the main reservoir of   nutrients for higher organisms. Most bacteria can  move on their own, and the majority of motile bacteria are  able to swim in viscous fluids using slender helical appendages called flagella. Low-Reynolds-number hydrodynamics is at the heart of the ability of flagella to generate propulsion at the micron scale. In fact, fluid dynamic forces  impact many aspects of bacteriology, ranging from the ability of cells  to reorient and search their surroundings  to their interactions within mechanically and chemically-complex environments. Using hydrodynamics as an organizing  framework, we  review the biomechanics   of bacterial motility and  look ahead to future challenges.
\end{abstract}
\maketitle

\section{Introduction}
Bacteria constitute the bulk of the biomass of our  planet. However, since we need a microscope to see them, we often forget  their presence. Observing life in the ocean, we see fish and crustaceans, but miss the marine bacteria that outnumber them many times over. On land, we see animals  and plants, but  often forget  that  a human body contains many more bacteria   than  mammalian  cells. While they are responsible for many infectious diseases, bacteria  play a critical role in the life of soils and higher organisms (including humans)   by performing chemical reactions and providing nutrients \citep{brock}. 

Ever since microscopes were invented, scientists  have worked to decipher the rules dictating the behavior of bacteria. In addition to quantifying the manner in which bacteria shape our lives and teach us about our evolutionary history, biologists  have long recognized that the behavior of bacteria is influenced by the physical constraints of their habitat, and by their evolution to maximize fitness under these constraints.  The most prominent of these  constraints is the presence of a surrounding fluid \citep{vogel96}.  Not only are all bacteria found in fluids, but being typically less than a hundredth  of a millimeter in length,  they experience much  larger viscous forces than inertial forces, and have adapted their behavior in response to these forces.  

Most   bacteria  are motile \citep{JarrellMcBride2008,kearns10} and this  motility is essential for the effectiveness of many pathogens \citep{ottemann97}.  The most common form of bacterial motility is swimming. In order to swim,  bacteria have evolved  flagella (originally from  secretory systems),    which are slender helical  appendages   rotated by specialized motors, and whose motions in viscous  environments  propel the cells forward  \citep{braybook}. The surrounding fluid can be seen both as a constraint and an advantage. It is the presence of the fluid itself and its interactions with three-dimensional bacterial flagellar filaments which allows cells to move and sample their chemical environment  --  a crucial step  for the cells to display robust and adaptive chemotactic responses to both  attractants and repellents, for nutrients, but also temperature, pH, and  viscosity \citep{purcell77,Berg1993,blair95,bergbook}.

But as they self-propel, bacteria are subject to the external constraints set by the physical world, and in particular by hydrodynamics. In this review, we highlight the consequences of fluid dynamics   relevant to the swimming of bacteria in viscous environments. Naturally, the biophysics of bacteria locomotion involves many aspects of soft matter physics, including  nonlinear elasticity, screened electrostatics, and biochemical noise, and hence fluid dynamics represents  but one feature of the complex balance of forces dictating the behavior of cells as they search their environment. The choice made here is to  use hydrodynamics as an organizing framework to overview various aspects of cellular locomotion.

We   start by  a biological overview of bacteria as cells, their geometry, the way they swim, and the variations among different species (\S\ref{sec:overview}). We  then review the basic  hydrodynamics features of flagellar propulsion (\S\ref{sec:hydrodynamics}) followed by the flows induced by swimming bacteria (\S\ref{sec:flow}). We  next address  the  flagellar polymorphs, their comparative hydrodynamics, and the potential relevance to evolution of the flagellum (\S\ref{sec:polymorphism}). With individual flagella understood, we  detail how fluid forces may play a role in the actuation of multiple flagella and be used  by cells to reorient (\S\ref{sec:reorientations}). The last three sections are devoted to locomotion near surfaces (\S\ref{sec:surfaces}), in external flows (\S\ref{sec:inflows}) and in complex fluids  (\S\ref{sec:complex}).

\begin{figure*}[t!]
\begin{center}
\includegraphics[width=0.8\textwidth]{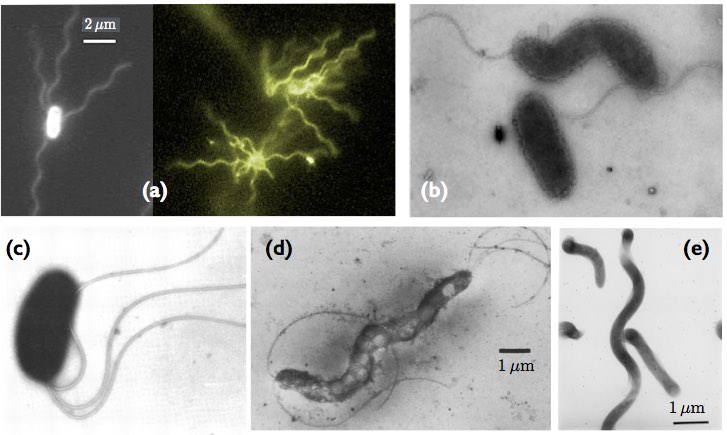}
\caption{Atlas of flagellated bacteria. 
(a) Peritrichous {\it Escherichia coli} (Howard Berg, Harvard University);
(b) Polar monotrichous {\it Pseudomonas aeruginosa} \citep{fsa08}; 
(c) Polar lophotrichous {\it Photobacterium fischeri} \citep{allen71}; 
(d) Polar amphitrichous  {\it Ectothiorhodospira hatochtoris} \citep{imhoff77}; 
(e) Spirochetes with endoflagella {\it Borrelia burgdorferi} \citep{goldstein96}. 
All figures reproduced with permission.
}
\label{fig1}
\end{center}
\end{figure*}

Throughout this review, the focus will be on single-cell behavior. Our purpose is to understand the way a single bacterium exploits, and is subject to, the physical constraints from the surrounding fluid. Other useful reviews can be used as complementary reading, in particular those highlighting the hydrodynamics of low-Reynolds number swimming \citep{lighthill75,brennen77,lp09}, the fluid dynamics of plankton  \citep{pedley92,stocker,goldstein14}, reproduction \citep{fauci06,gaffney11},  and collective cell locomotion 
 \citep{koch}.

\section{Biological atlas}
\label{sec:overview}

Bacteria come in many different shapes \citep{brock}. Those able to swim in fluids can be roughly divided into two categories: bacteria whose propulsion is driven by  helical flagellar filaments located outside a non-deforming cell body (the overwhelming majority, illustrated in Fig.~\ref{fig1}a-d) and those with spiral-like body  undergoing time-varying deformation (Fig.~\ref{fig1}e). 
 
\begin{figure*}[t!]
\begin{center}
\includegraphics[width=0.75\textwidth]{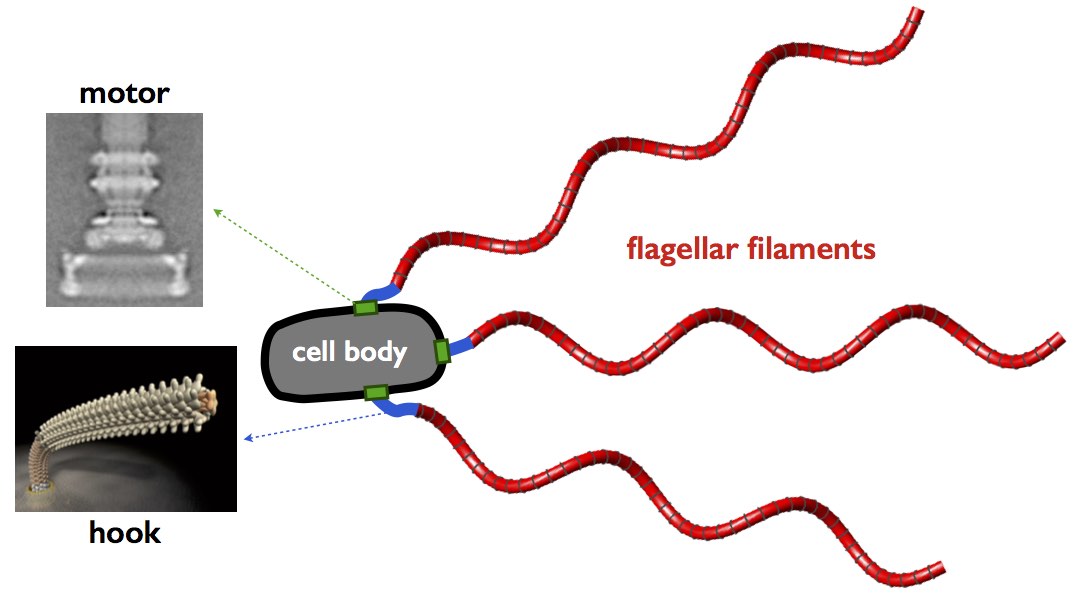}
\caption{Schematic representation of  a peritrichous bacterium with three flagellar filaments showing a bacterial rotary motor embedded in the cell wall (the stator part of the motor is not shown) \citep{berg03} and a hook with its junction to a flagellar filament (K.~Namba, Osaka University). All figures reproduced with permission.
}
\label{fig2}
\end{center}
\end{figure*}

Our understanding of how flagellated bacteria swim has its roots in  a series of  investigations in the 1970s showing conclusively  that -- unlike the active flagella of eukaryotes, which are active and muscle-like \citep{braybook} -- bacterial flagellar filaments  are passive organelles  \citep{BergAnderson1973,silverman74}.  Typically a few to ten microns in length, and only 40~nm in diameter, bacterial flagellar filaments are helical polymers  whose structure is discussed in detail in \S\ref{sec:polymorphism}. Each filament is  attached to a short flexible hook ($\approx$~60~nm in length, see Fig.~\ref{fig2})  acting as a universal joint. The hook is turned by a stepper motor driven by ion fluxes (bacterial rotary motor, Fig.~\ref{fig2}) which is well characterized  molecularly  \citep{berg03}.

Some flagellated cells are called {\it peritrichous} and have many motors (and thus many flagella) located randomly at various positions  on the cell body   (Fig.~\ref{fig1}a).  This is the case for the bacterium {\it Escherichia coli}  (or {\it E.~coli}), chosen as a model organism to study the fundamental biophysical aspects of cell motility and chemotaxis \citep{bergbook}. In contrast, {\it polar} bacteria have motors only located at poles of the  body. Of the polar bacteria, {\it  monotrichous} bacteria only have one motor, and thus one filament  (Fig.~\ref{fig1}b), {\it  lophotrichous} cells have a tuft of flagella originating from a single pole (Fig.~\ref{fig1}c)
while {\it amphitrichous} bacteria have flagella at each pole  (Fig.~\ref{fig1}d). Note that variations in this classification exist; for example, some monotrichous or lophotrichous bacteria are not polar. In all cases, the rotation of the motor is transmitted to the hook, which in turn transmits it to the helical filament, leading to the generation of hydrodynamic propulsive forces, as explained  in detail in \S\ref{sec:hydrodynamics}. 

Fluid dynamics was used to determine the torque-frequency relationship for the  rotary motor  by linking tethered  bacterial flagella to  beads of different sizes   in fluids of varying viscosity \citep{chen00}. The resulting  change in the Stokes resistance to bead rotation allowed to show that  the rotary  motor works at constant torque for a wide range of frequencies in the  counter-clockwise (CCW) directions (the frequency range depends strongly on temperature) before showing a linear decrease of the torque at higher frequencies \citep{berg03}. In contrast, when rotating in the clockwise (CW) direction, the motor torque linearly decreases with frequency \cite{yuan2010asymmetry}.

In the range relevant to locomotion, the motor can be assumed to behave as continuously rotating and producing a constant torque. The value of this torque is however still under debate and measurements are at odds with theoretical predictions. While its stall torque  has been experimentally estimated to be on the order of  $4,600$~pN~nm  \citep{berry97}, experiments using beads   lead to the conclusion that the motor has to be on the range $1260\pm 190$~pN~nm \citep{reid06}.  
In contrast, theoretical estimates using a simplified model for the fluid predict a significantly smaller value of  $370 \pm 100$~pN~nm   \citep{darnton07_torque}.

\begin{figure*}[t!]
\begin{center}
\includegraphics[width=0.65\textwidth]{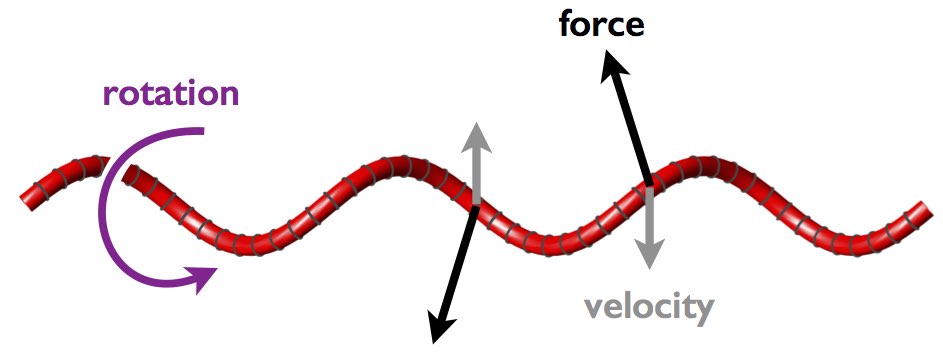}
\caption{Hydrodynamics of propulsion by a rotating helical flagellar filament. A left-handed helix rotates in the  CCW direction (when viewed from the right) as indicated by a purple arrow. The local velocity of two material points on the helix relative to the background fluid are illustrated with gray arrows. Since the fluid drag is larger (for a given velocity) for motion of the local helix perpendicular than parallel to the local tangent, the net force density from the fluid includes a component directed everywhere along the helix axis (black arrows). A net moment opposing the rotation is also generated, resulting in   counter-rotation of the cell.}
\label{fig3}
\end{center}
\end{figure*}

While the majority of swimming bacteria are powered by flagellar filaments rotating in the fluid outside the cells, two other modes of locomotion in fluids are notable.   
Bacteria of the spirochete family  have a spiral shape and undergo whole-body undulations powered also by  flagella  (Fig.~\ref{fig1}e). However,  in this case, the  filaments are constrained in the small space  between the cytoplasmic membrane and outer membrane of the cell body, and it is the rotation in this tight space which leads to wave-like whole-body undulations  and  in turn leads to locomotion in a fluid  \citep{vig12}. Swimming without flagella has also been reported for {\it Spiroplasma}, a type of helical bacterium whose body has no  cell wall.  
In that case, a wavelike  motion is  induced by  contraction of the cell cytoskeleton leading to the propagation of shape kinks along the body  \citep{ShaevitzLeeFletcher2005,wada07}.
 At least one other bacterium, the marine  {\it Synechococcus},  swims in fluids in  manner which is  as yet undetermined  \citep{ehlers12}.

To close this biological overview, we mention the other mode of fluid-based motility termed swarming \citep{JarrellMcBride2008,kearns10}.  A few bacterial families, when present in  media-rich in nutrients, can undergo a differentiation to a swarming state, where they become more elongated, grow more flagella and swim  in closely-packed groups, the physics of which is only beginning to be unraveled   \citep{copeland09,darnton10}.

\section{Hydrodynamics of helical propulsion}
\label{sec:hydrodynamics}

The most fundamental aspect of bacterial hydrodynamics is the ability of rotating flagella to produce propulsive forces \citep{chwang71,lighthill76,higdon79b}. At the origin of this generation of forces is  the non-isotropic nature of hydrodynamic friction in the Stokesian regime  \citep{brennen77,lp09}.  At the low Reynolds numbers relevant to swimming  bacteria (typically ranging from $10^{-4}$ to $10^{-2}$), the flows are governed  by the incompressible Stokes equations, and drag forces on rigid bodies scale linearly with their instantaneous velocities relative to the background fluid.   Because flagellar filaments  are slender (their aspect ratio is at least 100), their physics can be intuitively understood by analogy with the motion of a rod. At a given velocity, a  rod experiences a drag from the fluid about twice as large when translating perpendicular to its long axis as that  when moving along it. For a non-symmetric velocity/rod orientation, the hydrodynamic drag on the rod is thus not aligned with the direction of its velocity but possesses a nonzero component perpendicular to it -- hence the  possibility of drag-based thrust.

The extension of this  idea to case of a helical filament is illustrated in Fig.~\ref{fig3}. The flagellar filament is taken here to be left-handed and undergoes CCW rotation when viewed from the right looking left (more on this in \S\ref{sec:polymorphism}). The local velocity of two material points on the helix is illustrated by gray arrows: points in the foreground go down while points in the background go up. The instantaneous drag force density resisting the rotation of the helix in the fluid is  illustrated by black arrows. Because of  the higher resistance from the projection of the local velocity in the direction perpendicular to the local tangent to the helix, the total drag includes a nonzero component along the axis of the helix, which in both cases points to the left along the helix axis (this is exaggerated in the figure for illustration purposes). Consequently, a rotating helix  is subject to a net viscous force  aligned with the helix axis with an  orientation which  depends on the handedness of the helix and the direction of rotation.

From a Stokes flow standpoint, the grand resistance matrix for a helical flagellar filament,  which linearly relates forces and moments to velocities and rotation rates, includes therefore an off-diagonal term coupling axial rotation  with axial force \citep{kimbook,purcell97}. Because the whole cell is force-free, it has to swim in order to balance this force. 
This is the physical mechanism used by bacteria to self-propel in fluids, and it  will in fact be intuitive to anybody having opened a bottle of wine with a corkscrew where the rotation of the screw induces  its translation inside the cork.  As a difference with a corkscrew however, the propulsion of a helix in a viscous fluid produced by its rotation is very inefficient, both kinematically (small forward motion per rotation) and energetically (only a few percent of the total work done by the helix   is transmitted to propulsive work)  \citep{purcell97,chattopadhyay06,spagnolie_helix}. 
The same hydrodynamic principles  can be exploited to design artificial helical swimmers powered by rotating magnetic fields   \citep{Ghosh2009,Zhang2010b}.  A famous illustration of helical locomotion in fluids is provided in G.I. Taylor's  movie on low Reynolds number flows \cite{taylor67}.

In addition to  forces, the rotating filament in Fig.~\ref{fig3} is subject to a net hydrodynamic moment resisting the rotation. In order for the bacterium to remain torque-free, the cell body needs therefore to counter-rotate. With  flagella typically rotating at $\approx 100~{\rm Hz}$, this is achieved by a  counter-rotation 
at smaller frequency $\approx 10~{\rm Hz}$ due to small rotational mobility of a large cell body \citep{chwang71,Magariyama_etal1995}.  Because of this balance of moments, locomotion powered by helical flagella is not possible in the absence of a cell body, a result qualitatively different from locomotion induced by planar waving deformation, for which a head is not required \citep{lp09}. The resulting motion of a bacterium in a fluid  consists thus of flagellar filaments rotating, the cell body counter-rotating, and the whole  cell moving forward approximately in a straight line, or in  helices of small amplitude when the axis of the filaments is not aligned with that of the cell body \citep{keller76,hyon12}. In general the cell body does not induce additional propulsion, which is all due to the flagella, unless it is the right shape and has the right orientation relative to the helical flagellum  \citep{chwang1972locomotion,liu14}.

The hydrodynamic theories developed to calculate  the  distribution of forces on rotating helical flagella are of three flavors \citep{brennen77,lp09}. The first one, historically,  considered the theoretical limit of very (exponentially) slender filaments to compute forces as a perturbation expansion in the logarithm of the aspect ratio of the flagella. At leading order, this so-called resistive-force theory predicts that the local force density at one point along the filament  is proportional to its  local velocity relative to the fluid with a coefficient of proportionality which depends on the relative orientation of the velocity with respect to the local tangent \citep{gray55,cox70,lighthill75}.  This is therefore the natural extension of the physical picture discussed above for rigid rods to deforming filaments. While this is  the most widely used approach, and one which is is able to provide  basic physical pictures, it is often not sufficiently accurate as the associated relative errors are only logarithmically small  \citep{johnson79,chattopadhyay09}. More precise is the second approach, termed slender-body theory,  which consists in distributing appropriate flow singularities of suitable strengths in order to match the boundary condition on the surface of the filament at a required, algebraically small, precision    \citep{lighthill76,hancock53,johnson80}. In that case the equation relating the distribution of velocity to the distribution of forces is nonlocal, because of long-range hydrodynamic interactions, and thus needs to be inverted numerically. A final option is to use fully three-dimensional computations.  The most common method is an  integration of the equations of Stokes flows using either boundary elements \citep{phan87,liu13}, regularized flow singularities \citep{flores05,cortez13}, or  immersed boundaries \citep{fauci06}. Mesoscopic particle-based methods have recently been proposed as an alternative approach \citep{reigh12}.

\section{Flows induced by bacteria}
\label{sec:flow}

\begin{figure*}[t!]
\begin{center}
\includegraphics[width=0.8\textwidth]{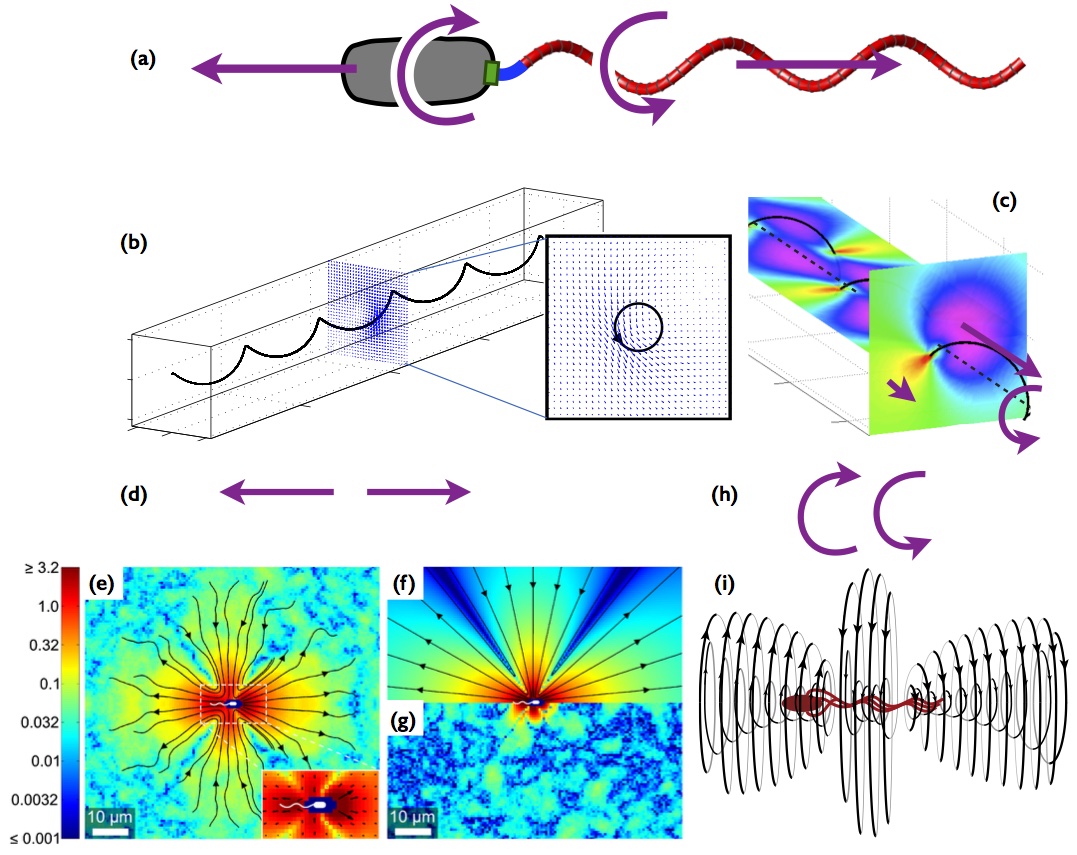}
\caption{The flows produced by swimming bacteria.
(a) Schematic representation of the forcing induced locally by the bacterium on the surrounding fluid (purple arrows); 
(b) Rotational flow near a rotating helical filament \citep{spagnolie_helix}; 
(c) Axial flow   along the axis  of the helical filament, directed away from the cell body; 
(d) In the far-field, the cell acts on the fluid as a force dipole; 
(e) Measurement of the flow induced by swimming {\it E.~coli}  \citep{drescher11}; 
(f) Theoretical prediction from force dipole; 
(g) Difference between measurement and theoretical  prediction; 
(h) The rotation  of the helical filament  and the counter rotation of the cell body acts on the fluid as a rotlet dipole; 
(i) Simulation of this rotational flow  with streamlines  \citep{watari10}.
All figures reproduced with permission.
}
\label{fig4}
\end{center}
\end{figure*}

The general fluid dynamics picture of a swimming bacterium powered by the rotation of a flagellum is illustrated in Fig.~\ref{fig4}a. We use purple arrows to indicate the local forcing of the swimming cell on the surrounding fluid.  Near the filament, the induced flow  is a combination of rotation and polar axial pumping  directed away from the cell body. This is illustrated  in Fig.~\ref{fig4}b-c using slender-body theory for a helical waveform of diameter 400 nm and wavelength 2.3~$\mu$m (the so-called ``normal'' waveform, see discussion in \S\ref{sec:polymorphism}) \citep{spagnolie_helix}.

Looking further away from the cell, because the swimming bacterium is force-free,  it pushes on the fluid with an equal and opposite force to the one generated by the rotating filament  (Fig.~\ref{fig4}a), and thus the leading order flow in the far-field is a force-dipole, or stresslet, as schematically illustrated in Fig.~\ref{fig4}c  \citep{Batchelor1970,ishikawa07_bacteria}. The type of associated dipole is termed a pusher, in contrast with puller dipoles relevant to   cells which swim flagella first, for example planktonic bi-flagellates  \citep{stocker}. 
This flow has been  measured for swimming {\it E.~coli} using an experimental tour-de-force  \citep{drescher11}. The magnitude of the induced flow is shown in Fig.~\ref{fig4}e (in $\mu$m/s), while the theoretical dipolar prediction is shown in Fig.~\ref{fig4}f, with a difference shown in Fig.~\ref{fig4}g. The dipole strength is on the order of   $0.1-1$ N~$\mu$m \citep{drescher11,berke08} and this  dipolar picture is very accurate up to a few  body lengths away,  at which point the flow speeds get overwhelmed  by thermal noise.

While the leading-order flow is a stresslet decaying spatially as $1/r^2$, the velocity field around a bacterium  also includes   higher-order (decaying as $1/r^3$) flow singularities which are important in the near field. These include source-dipoles and force quadrupoles \citep{chwang75}. Force-quadrupoles, for example, dominate velocity correlations between swimming bacteria, because force-dipoles are front-back symmetric \citep{liao07}. An  important flow near a  swimming  bacterium is a rotlet-dipole. Since the flagellum and the cell body rotate relative to the fluid in opposite directions (Fig.~\ref{fig4}a), they act on the fluid as two point moments (or rotlets) of equal and opposite strengths, hence a rotlet-dipole, decaying spatially as $1/r^3$, and shown schematically in Fig.~\ref{fig4}h (a rotlet dipole is a particular force quadrupole). A computational illustration of the streamlines associated with this flow is shown in Fig.~\ref{fig4}i \citep{watari10}. Note that  both the force-dipole and the rotlet-dipole can be time-dependent, and in some cases induce  instantaneous flows  stronger than their time averages \citep{watari10}.

The flows induced by swimming cells have been useful in understanding  two   problems. First, flows created by populations of  bacteria  lead to enhanced transport in the fluid. In addition to  Brownian motion, passive tracers (for example nutrient molecules, or much larger suspended particles) also feel the sum of all hydrodynamic flows from the population of  swimming bacteria, and in general will thus display an increase in the rate at which they are transported.
Enhanced diffusion by swimming bacteria was first reported experimentally for flagellated {\it E.~coli} cells with non-Brownian particles  \citep{wu00}, followed by work on molecular dyes \citep{kim04_diffusion}. Dilute theories and simulations based on binary collisions between swimmers (possibly with a finite persistence time or finite run length) and passive tracers (possibly Brownian) have allowed to predict diffusion constants 
\citep{Lin11,mino13,pushkin13jfm,kasyap14}. In particular, it was shown that the extra diffusivity remains linear with the concentration of bacteria well beyond the dilute regime. The effect, while negligible  for  small molecules with high Brownian  diffusivity, can be significant for large molecules or non-Brownian tracers.

The second topic where swimming-induced flows are critical is that of collective swimming. Suspension of swimming bacteria display  modes of locomotion qualitatively different from individual cells, with long orientation and velocity correlation lengths, instabilities, and an intrinsic randomness -- a process recently referred to as  {\it bacterial turbulence}.  Building on the standard theoretical description for Brownian suspensions  \citep{doi88},  the  modeling approach consists in describing the populations of swimming bacteria as continuua using a probability density function in orientation and position. Conservation of probability then establishes a balance between changes in bacteria configurations and diffusion, with  changes in both position and orientation which are brought about by the flows (and their gradients) induced by the swimming bacteria. While the current  article is focused on single-cell behaviour, we refer to the recent Annual Review on the topic   \citep{koch} and book \citep{CFBS2015}, and references therein, for a more detailed overview.

\begin{figure*}[t!]
\begin{center}
\includegraphics[width=0.98\textwidth]{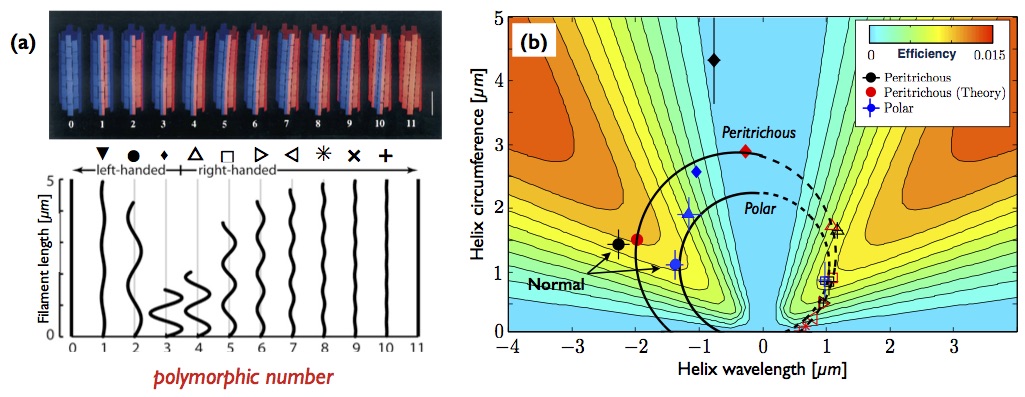}
\caption{
(a) Illustration of the  twelve polymorphic shapes of bacterial flagella, numbered  \#0 to \#11 \citep{hyn98,darnton07_torque}; 
(b) The intrinsic hydrodynamic efficiencies of each  polymorph for
Calladine's theory (red) and for 
experimentally-measured  peritrichous  (black)  and polar (blue) flagellar polymorphs \citep{purcell97,spagnolie_helix}. Left (resp.~right)-handed helices are denoted with negative (resp.~positive) wavelengths and the corresponding waveform with filled (resp.~empty) symbols, as seen in Fig.~\ref{fig5}a. 
The circles allow differentiating peritrichous from polar polymorphs. 
All figures reproduced with permission.
 }
\label{fig5}
\end{center}
\end{figure*}

\section{Polymorphism and the evolutionary role of hydrodynamics}
\label{sec:polymorphism}

It is one of nature's wonders that bacterial flagellar filaments only exist in discrete helical  shapes called flagellar polymorphs  \citep{hyn98,Calladine78,namba97}. Each filament is a polymer composed of a single protein (flagellin) which  assembles in long protofilaments. Eleven of these protofilaments wrap around the circumference of the filament. Because flagellin (and thus each protofilament) exists in two distinct  conformation states,  an integer  number of different conformations is available for the flagellum, of which only twelve are molecularly and mechanically stable \citep{Calladine78,srigiriraju06}. These are  the helical ``polymorphic'' shapes, illustrated in Fig.~\ref{fig5}a. Of these 12 shapes, 2 are straight, and 10 are true helices. Of these 10 helices, 9 have been observed experimentally, induced by chemical, mechanical  or temperature modifications \citep{Leifson60,Hotani80,hotani1982}. Of the 10 non-straight shapes, 3 are left-handed -- including the most common one, called ``normal,'' whose polymorphic number is \#2 --  and 7 are right handed -- including the important ``semi-coiled,'' which is \#4, and  ``curly,'' which is \#5. 

As an elastic body, the  force-extension curve of individual flagellar filaments  and the transitions between the different polymorphs have been precisely characterized experimentally using optical tweezers pulling on filaments  attached to a glass surface \citep{darnton07}. Fitting the data to  a Kirchoff elastic rod model with multiple minima for curvature and twist  allows then  to determine the elastic constants around polymorphic minima. The bending rigidities so estimated are on the order of few pN~$\mu$m$^2$, and the transition forces from one shape to the next are on the order of a few pN  \citep{darnton07}. A variety of associated modeling approaches for flagellar mechanics  have been developed to reproduce these experimental results \citep{wada08,stark10}.  That  flagellar filaments are elastic means they might deform under rotation, but during normal swimming conditions this has been shown to be  a small effect \citep{kim_powers2005}.  
It is however possible that the moments and forces from the fluid induce buckling of a rotating filament, similar to elastic  rods  buckling under their own weight. While the threshold rotation rates appear to be close to biological numbers \citep{vogel12}, no sign of flagellar buckling has been reported so far.

\begin{figure*}[t!]
\begin{center}
\includegraphics[width=0.8\textwidth]{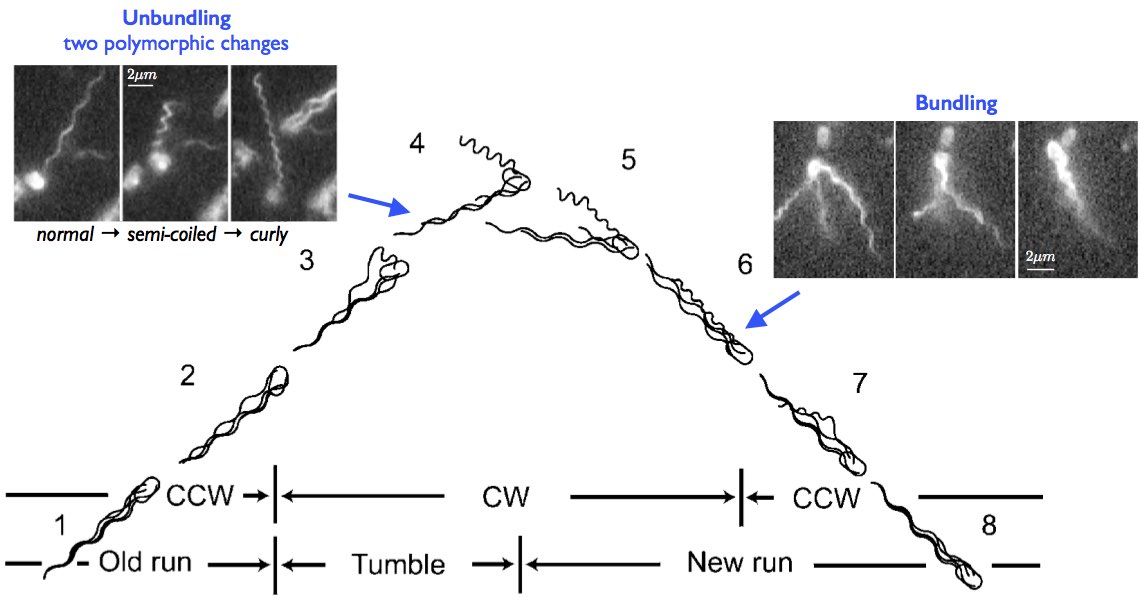}
\caption{A run-to-tumble-to-run transition for swimming {\it E.~coli} as a model for the reorientation mechanism of peritrichous bacteria 
\citep{turner05,darnton07_torque}. During a run all motors turn CCW while in a tumble at least one motor turns CW which induces polymorphic changes in the associated flagellar filament, typically from normal to semi-coiled and then curly (unbundling) before returning to the normal shape at the beginning of a new run (bundling). All figures reproduced with permission.
}
\label{fig6}
\end{center}
\end{figure*}

The fact that different shapes exist is important for locomotion because, as we see in the next section, when a bacterium turns, the  direction of rotation of one (or more) bacterial motor is changing, which induces a rapid  polymorphic transformation for the associated filament, and in particular a flip of its chirality. This was modeled  in a biophysical study for the bacterium {\it Rhodobacter sphaeroides} with  two different polymorphs \citep{vogel13}. Similar  chirality transformations have been obtained in the lab using tethered flagella in external flows \citep{hotani1982,coombs02}. A  helical flagellum held in a uniform stream is subject to a flow-induced tension as well as hydrodynamic moments, which are able to induce  chirality reversal and cyclic transformations between right (curly or semi-coiled) and left-handed (normal) helices. 

One aspect of polymorphism where fluid dynamics plays a fascinating role is the propulsive performance of each flagellar waveforms. For each waveform, an intrinsic propulsive efficiency may be defined from the mobility matrix  of the helix, comparing the power used for locomotion to the  power expanded by the rotary motor against the fluid \citep{purcell97}. Physically, efficient helices are those which maximize the  propulsive coupling between rotation and translation while minimizing the resistance in both translation and rotation.  A near-rod helix, such as polymorph \#9 
in Fig.~\ref{fig5}a, or one with a  small  wavelength,  such as polymorph \#3, will be very inefficient. Which one of the different shapes is the most efficient? This is illustrated in  Fig.~\ref{fig5}b which  plots the isovalues of the intrinsic hydrodynamic efficiency as a function of the helix wavelength and the circumference of the cylinder on which it is coiled \citep{spagnolie_helix}. Left-handed (resp.~right-handed)  helices are denoted with negative (resp.~positive) wavelengths. Superimposed on the color map are the geometrical measurements from the actual polymorphs for peritrichous (black symbols) and polar flagella  (blue symbols)  together with the theoretical shapes predicted by molecular theories (red symbols) \citep{Calladine78}. Filled (resp.~empty) symbols refer to the left-handed (resp.~right-handed) polymorphs  shown in Fig.~\ref{fig5}a. The two  circles allow one to distinguish between peritrichous and polar polymorphs. In all cases, the most efficient waveform is the left-handed normal polymorph (\#2 in Fig.~\ref{fig5}a). This is the waveform displayed by bacteria  during forward locomotion. Further, not only is the normal helix the most efficient among all  polymorphs, it is in fact close to being the overall optimal (which is at the intersection of the black circles and the warm color in the map). In addition, further analysis of the results reveals that the second and third most efficient waveforms are the right-handed semi-coiled (\#4 in Fig.~\ref{fig5}a) and curly (\#5), both of which are the waveforms used by wild-type swimming bacteria to change directions (see \S\ref{sec:reorientations}).  These results suggest that fluid dynamics and mechanical efficiencies  might have provided strong physical constraints in the evolution of bacterial flagella.


\section{Hydrodynamics of reorientations}
\label{sec:reorientations}

Swimming is the mechanism which allows bacteria to sample their environment \citep{purcell77}. Because they are small, bacteria are not able to sustain directional swimming for  extended periods of time, and as a consequence they always display diffusive behavior at long times. For swimming at speed $U$ for a time $\tau$ followed by a random change of direction, the long-time behavior is diffusive, $r^2 \sim D t$ with the diffusivity scaling as $D\sim U^2 \tau$, typically much larger than  Brownian  diffusivities \citep{Berg1993}.  When this diffusive dynamics  is biased using  sensing from the cells of  their chemical environment, it leads to net drift and {chemotaxis} \citep{schnitzer93,bergbook}.

While the value of $U$ is dictated by the hydrodynamics of swimming, what sets the time scale $\tau$ and how do cells manage to reorient? The time scale is set internally by the sensing apparatus of bacteria, a complex but well-characterized internal protein network  which chemically  links the information gathered by receptors on the cell membrane to the action of the rotary motor \citep{falke97}. In order for the cells to then reorient, three  mechanisms have been discovered, each taking advantage of different physics, and each involving fluid dynamics \citep{mitchell02}.

The first one is termed run-and-stop, where a bacterium swims and then stops, simply waiting for thermal noise to reorient the cell. The time scale is therefore set in this case by a balance between the magnitude of thermal noise and the viscous mobility of the whole cell in rotation, leading to reorientation times on the order of seconds. The second mechanism, termed run-and-reverse, is used by many  polar  marine bacteria \citep{stocker12}. In that case, the rotary motor alternates between CCW and CW rotations. Kinematic reversibility for the surrounding fluid would predict that  by doing so the  cells should keep retracing their steps. However,  the propulsive forcing from the fluid on the flagellar filament, which is transmitted to the flexible  hook, is able to  induce buckling of the  hook  which leads to a random reorientation of the cell, thereby escaping the reversibility argument  \citep{xie11,son2013bacteria}.

The third, and most-studied, mechanism is called run-and-tumble, used by peritrichous bacteria such as {\it E.~coli}, and whose details were made clear by a  pioneering  experiment allowing to visualize flagellar filaments in real time \citep{turner05}  (see illustration in Fig.~\ref{fig6}). Critical to the effective locomotion in this case is the ability of multiple flagellar filaments  to bundle when they all rotate in the same direction, and unbundle otherwise \citep{macnab1977,magariyama01}.   When a bacterium runs, all its flagellar filamenta adopt the normal waveform,  turn in a CCW fashion (when viewed from  behind the cell looking in), and are gathered in a tight, synchronized helical  bundle behind the cell.  Based on chemical clues from its environment,  the cell might then initiate    a tumbling event (which is Poisson distributed), for which at least one motor (but possibly more)  switches its direction rotation to CW,  inducing a polymorphic change of  its flagellar filament  from normal to semi-coiled (in most cases) which at the same time flies out of the bundle, resulting in a change of orientation for the cell body. As the motor continues to rotate CW, the flagellum switches to the curly shape, and then when the motor reverts back to a CCW rotation, comes back to the normal shape and returns into the bundle. These short tumbles typically last $\tau\sim 0.1s$ while the runs last about one second. Note that many mutants of {\it E. coli} exist, either naturally occurring or genetically induced in the lab, displaying some variation in geometry or behavior; for example, the mutant HCB-437 does not tumble and is smooth swimming \cite{wolfe1988acetyladenylate}.

The bundling and unbundling of flagellar filaments compose a large-deformation fluid-structure interaction problem at low Reynolds numbers combined with short-range flagella-flagella electrostatic repulsion \citep{weibull50,lytle02}.  Due to this complexity, physical studies of bundling have taken two drastically different approaches. On one hand, investigations have been carried out on very simplified geometries and setups, either numerically   \citep{kim04,reiehert05} or experimentally   at the macroscopic scale \citep{macnab1977,kim03,Qian_etal2009}, in order to probe synchronization between driven identical helices and their bundling. It was found that nonlocal hydrodynamic interactions between helices were   sufficient to induce attraction, wrapping, and synchronization  provided the helices were of the  right combination of chirality (both left-handed and rotating CCW or their mirror-image equivalent) but some amount of elastic compliance was required.  Biologically, it was argued that this flexibility could be provided by the hook.

\begin{figure*}[t!]
\begin{center}
\includegraphics[width=0.8\textwidth]{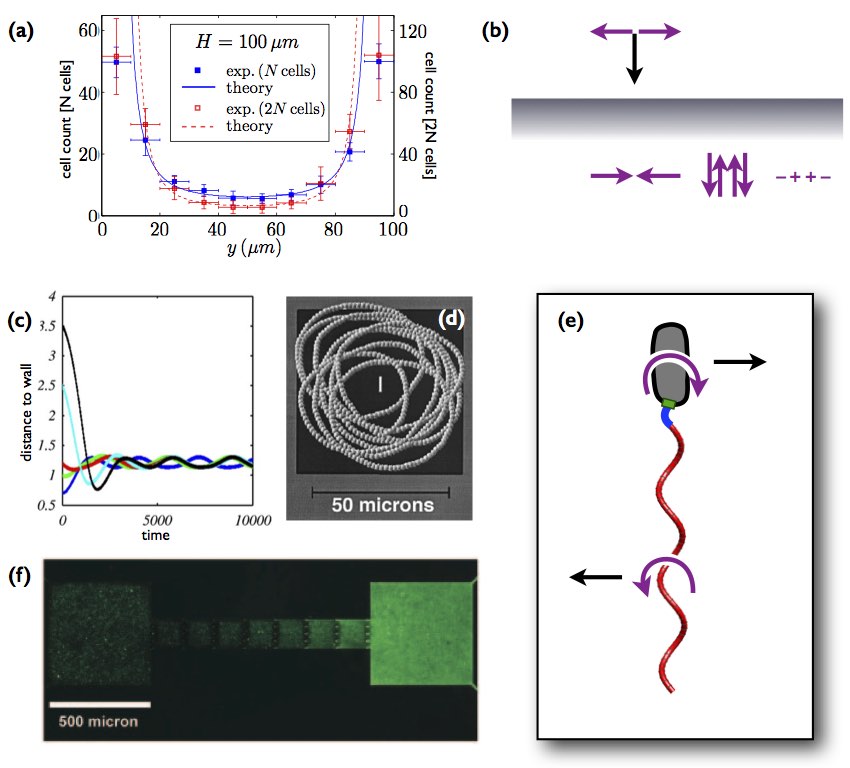}
\caption{
 (a) Attraction of smooth-swimming {\it E.~coli} cells by two rigid surfaces located at $y=0$ and $y=100$~$\mu$m (symbols) and fit by a point-dipole model (lines)  \citep{berke08}; 
 (b) Hydrodynamic images  for a pusher force dipole above a flat no-slip surface  (purple arrows) leading to  attraction  by the surface (black arrow); 
 (c)  Computations showing stable swimming at a finite distance from a rigid surface for a  polar bacterium (distance to wall nondimensionalized by cell size as a function of time nondimensionalized by flagellar rotational frequency)   \citep{giacche10}; 
 (d) Tracking data for a smooth-swimming {\it E.~coli} cell  near a glass surface  \citep{vigeant97}; 
 (e) Physical interpretation of wall-induced moment on  bacterium with helical flagellar filament leading to circular swimming: the rotating  flagellum is subject to a surface-induced force to its left (when viewed from above) and the counter-rotating body by a force in the opposite direction, leading to a  moment and therefore rotation of the cell;  
 (f) Rectification of fluorescent swimming {\it E.~coli} in channels with asymmetric geometrical features  \citep{funnels}. 
All figures reproduced with permission.
 }
\label{fig7}
\end{center}
\end{figure*}

On the other hand, some groups have attempted to tackle the complexity of the problem using realistic setups of deforming flagellar filaments from various (often formidable) computational angles,  including the use of regularized flow singularities \citep{flores05}, the immersed boundary method \citep{lim12}, 
mesoscale  methods \citep{reigh12},  boundary elements  \citep{kanehl14} and bead-spring models adapted from polymer physics \citep{watari10,graham11}. The numerical results confirm the physical picture of bundling and unbundling driven by a combination of flexibility and hydrodynamic interactions, with a critical dependence  on the  geometrical details, the relative configuration of the flagella, and the  value of the motor torques driving the rotation.


\section{Bacteria swimming near surfaces}
\label{sec:surfaces}

Surfaces are known to affect the behavior of bacteria \citep{vanloosdrecht90,harshey03}. One of the most striking aspect of locomotion near boundaries is the tendency of swimming bacteria to be attracted to surfaces, with  steady-state concentration showing 10 fold-increases  near surfaces  \citep{berke08}. This is illustrated in Fig.~\ref{fig7}a for smooth-swimming (i.e.~non-tumbling) {\it E.~coli} cells. The attraction mechanism was first argued to be purely hydrodynamic in nature \citep{hernandez-ortiz05,berke08}, easily understood as an image effect. As seen in \S\ref{sec:flow}, the flow perturbation induced by a swimming bacterium is (at leading order in the far field) a pusher force dipole. The hydrodynamic image system necessary to enforce a no-slip boundary condition on a flat surface includes a force dipole, a force quadrupole, and a source dipole, all located on the other side of the surface (purple arrows in Fig.~\ref{fig7}b) \citep{Blake1971}. The net effect of these image singularities on the original dipole is an hydrodynamic attraction toward the wall (black arrow). Intuitively, a pusher dipole draws fluid in from its side, and therefore  pulls itself  toward a surface when swimming parallel to it. Similarly, the hydrodynamic moment induced by the  image system is able to rotate pusher dipoles so that their stable orientation is swimming parallel to the surface \citep{berke08}, leading to kinematic asymmetries for bacteria switching between pusher and puller modes \citep{magariyama05}. 

Experimental measurements of the flow induced by bacteria showed, however, that this dipolar flow gets quickly overwhelmed by noise a few cell lengths away, and thus acts only when the cells are close to the surface, leading to long residence times   \citep{drescher11}. For cells which swim toward a surface and get trapped there,  what determines the long-time equilibrium distance  between swimming cells  and surfaces, and their orientation? Although simple models  all predict bacteria eventually touching  walls and therefore  require short-range repulsion \citep{dunstan2012two,spagnolie12}, boundary element computations with a  model polar  bacterium demonstrated that hydrodynamics alone was able to select a stable orbit with swimming at a finite distance (Fig.~\ref{fig7}c)  \citep{giacche10}.

A second remarkable feature of bacteria swimming near surface is a qualitative change in their trajectories. Between reorientation events, flagellated bacteria far from walls swim along straight or wiggly paths. Near rigid surfaces, their trajectories are observed to become circular, typically CW when viewed from above the surface in the fluid \citep{berg_1990,frymier95}.   This is illustrated in Fig.~\ref{fig7}d for an individually-tracked  {\it E~coli} cell swimming near a glass surface \citep{vigeant97}.  Early computations for model polar bacteria also predicted  circular trajectories \citep{ramia93} while  recent experiments
 showed that the physicochemical nature of the interface plays a critical role selecting the  rotation direction     \citep{lemelle_2013,morse_2013}.

\begin{figure*}[t!]
\begin{center}
\includegraphics[width=0.77\textwidth]{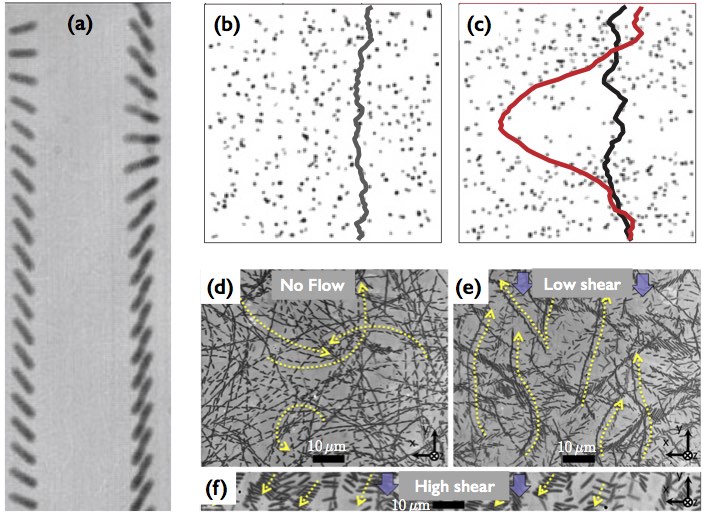}
\caption{
(a) Jeffery's orbits for non-flagellated {\it E.~coli} cells in a microchannel \citep{kaya09}; 
(b) Non-motile {\it Bacillus subtilis} cells show a uniform concentration  under non-uniform shear \citep{rusconi14};  
(c) Motile cells initially uniformly distributed (black line) accumulate under non-uniform shear  in the high-shear regions (red line); 
(d) Motile {\it E.~coli}  cells swimming in circles near rigid surfaces in the absence of flow  \citep{kaya12}; 
(e) Upstream swimming of the same cells under moderate shear; 
(f) Under strong shear, all cells are advected downstream. 
All figures reproduced with permission.
}
\label{fig8}
\end{center}
\end{figure*}

 The switch from forward to circular swimming is a fluid dynamics effect owing  to new forces and moments induced near surfaces  \citep{lauga06_circles}. This is illustrated in Fig.~\ref{fig7}e for locomotion of model polar bacterium above of a rigid wall.  A helix rotating parallel to a no-slip surface is subject to a net force whose direction depends on the chirality of the helix and its direction of rotation (to the left in Fig.~\ref{fig7}d for a CCW rotation of a left-handed helix). Similarly, as the cell body is counter-rotating, it is subject to a hydrodynamic force in the direction  of its rolling motion on the surface (to the  right in Fig.~\ref{fig7}d) \citep{kimbook}.  These two forces are exactly zero far from the surface and their magnitudes depend strongly on the distance to the wall  \citep{li08}. The net effect of these forces is a wall-induced moment acting on the cell. Since the cell is torque-free, it needs to rotate (here, CW when viewed from above), leading to circular trajectories. Notably, the sign of the surface-induced forces are reversed if the rigid wall is replaced by a free surface, leading to circular swimming in the opposite direction   \citep{dileonardo_2011}, and a possible co-existence between CW and CCW for cell populations near complex interfaces \citep{lemelle_2013,morse_2013,lopez14}.

The interactions of bacteria with complex geometries have also lead to fascinating results. Swimming {\it E.~coli} cells in rectangular  microchannels with three polymeric walls and one agar surface showed a strong 
 preference to swim near the agar side of the channels
  \citep{diluzio05}, a phenomenon still unexplained. The same bacteria are rectified by channels with asymmetric geometrical features \citep{funnels}, which can then be exploited to passively increase cell concentrations (Fig.~\ref{fig7}f). Similarly, bacteria can be harnessed  to perform mechanical work and, for example, rotate  asymmetric micro-gears   \citep{sokolov10,diLeonardo10}. In strong confinement, helical flagella continue to  induce  propulsion, but the swimming speed decreases for motion driven at constant torque \citep{liu14_2}. Related experiments show that the limit of  flagella-driven motility  is reached when the cell body has clearance of about less than 30\% of its size, although cells can continue to move in smaller channels through cell division \citep{mannik09}.

\section{Bacteria  in flows}
\label{sec:inflows}

Similarly to  colloidal particles, swimming bacteria respond  to external flows by being passively advected and rotated. In many relevant situations where the cells are much smaller than any of the flow length scales (e.g.~in marine environments), the flow can be assumed to be locally linear, and thus a superposition of uniform translation, pure extension, and pure rotation. In that case, the dynamics of the (elongated) bacteria is described by Jeffery's equation which is exact for prolate spheroids in linear flows \citep{pedley92}. Physically, the cells move with the mean flow velocity and their rotation rate is  the sum of the vorticity component (which alone would predict rotation at a constant  rate) and the (potential) extensional  component (which alone would predict steady alignment with the principal axis of extension).  The resulting tumbling trajectories, called Jeffery's orbits, are illustrated in Fig.~\ref{fig8}a for non flagellated {\it E.~coli} cells under shear in a microfluidic device \citep{kaya09}. 

In addition to the classical Jeffery's orbits, bacteria are subject to an additional mechanism which results from their chiral shapes.   The hydrodynamics of a  helical  flagellum in a simple  shear flow  leads to a viscous torque being produced on the helix, which reorients it away from the plane of shear. Combined with motility, this is an example of  {\it rheotaxis}  where cells responds to shear  (here, purely physically) \citep{marcos12}. Beyond the assumption of linear flow, swimming bacteria respond to non-uniform shears by developing  non-uniform orientation distributions, which, when coupled to motility, lead to cells accumulating in the high shear regions  \citep{rusconi14}. This is illustrated in  Fig.~\ref{fig8}b-c for {\it Bacillus subtilis}  in Poiseuille flow. Non-swimming cells have a uniform distribution (Fig.~\ref{fig8}b) whereas motile cells starting uniformly distributed (black line in Fig.~\ref{fig8}c) develop non-uniform distributions and gather in  the high-shear regions (red line Fig.~\ref{fig8}c).

\begin{figure*}[t!]
\begin{center}
\includegraphics[width=0.8\textwidth]{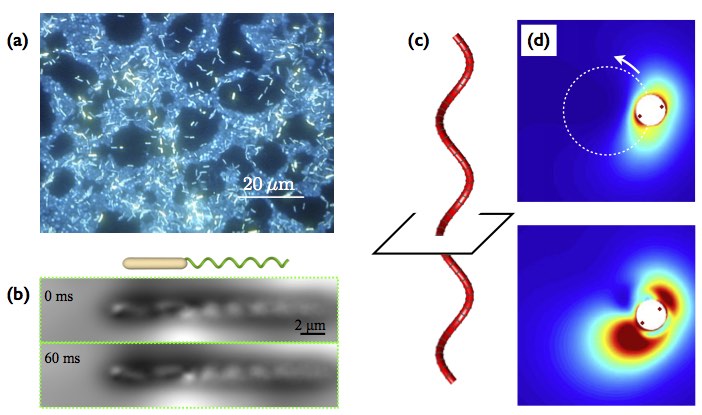}
\caption{
(a)  Biofilm on the  stainless steel surface  of a bioreactor  (Centers for Disease Control and Prevention);  (b)  Waves in the director field  created by rotating flagella of the peritrichous  {\it B. subtilis} in a liquid crystal \citep{zhou14};   
(c-d) Magnitude of the polymer stresses  in the  cross section of a flagellum rotating in a Oldroyd-B fluid  (nondimensionalized by the viscosity times the filament rotation rate)  from computations  (the white dashed line indicates the path followed by the flagellum  centerline)
\citep{spagnolie13}. Top: $De=0.3$; bottom: $De=1.5$. 
All figures reproduced with permission.
}
\label{fig9}
\end{center}
\end{figure*}

 One compelling aspect of the response of bacteria to external flows  is the  feedback of swimming cells   on the surrounding fluid through the stresses they exert.  Due to their dipolar flows, swimming bacteria exert bulk stresses on the fluid -- hence the term ``stresslet'' originally coined by Batchelor in the context of passive suspensions \citep{Batchelor1970}. A group  of swimming bacteria subject to an external deformation  will thus generically respond by a state of stress with a rich dynamics.  The most fundamental question relating stresses to deformation in the fluid is: what is the effective viscosity of a suspension of swimming bacteria? Experiments have been first carried out with the   peritrichous  
 {\it Bacillus subtilis}  by estimating the viscosity in two different ways, first by  looking at the unsteady decay of vortex and then  by measuring directly the torque on a rotating probe in the  cell suspension \citep{sokolov09}. The results showed a strong reduction in the effective viscosity (up to a factor of seven), while an increase was possible at larger cell concentrations. A different setup was proposed for  {\it E. coli} where a suspension was flown in a  microchannel and the  effective viscosity estimated using  predictions   for  the unidirectional flow profile   from multiphase Newtonian fluid dynamics \citep{gachelin13}. In that case, the effective viscosity showed an initial decrease, followed by  shear-thickening  up to relative viscosities above 1, followed by   shear-thinning at high shear rates. In parallel, theoretical and computational work on model cells in both shear and extensional flows predicted the viscosity decrease for pusher cells, together with the possibility of negative viscosities and   normal stress coefficients of sign opposite to the ones for passive suspensions \citep{haines09,saintillan_ext,saintillan_shear}.

Finally, bacteria   swimming near boundaries in external flows  also have a propensity to swim upstream and hence against the flow \citep{cisneros07,ecoli_upstream_yale}. This is illustrated in Fig.~\ref{fig8}d-f where motile {\it E.~coli}  cells which  swim in circles  in the absence of flow  (Fig.~\ref{fig8}d) are seen to  move upstream under moderate shear (Fig.~\ref{fig8}e) and are passively  advected downstream at high shear  (Fig.~\ref{fig8}f) \citep{kaya12}. The physical origin of this transition to upstream swimming has been proposed to be hydrodynamic, whereby  a shear flow  of moderate strength produces a torque on a downward-facing cell which reorients it so that its stable equilibrium is to point upstream and thus swim against the flow \citep{ecoli_upstream_yale,kaya12}.

\section{Locomotion in complex fluids}
\label{sec:complex}

Bacteria who inhabit soils and higher organisms routinely have to progress through complex  environments with  microstructures whose characteristic  length scales might be similar to that of the cell.   Some bacteria also have to endure chemically- and mechanically-heterogeneous environments.   For example,  {\it Helicobacter pylori}, which survives in the acidic environment of the stomach, creates  an enzyme which locally increases pH and can induce rheological changes to the mucus protecting the epithelium surface of the stomach, changing it from a gel to a viscous fluid in which it is able to swim \citep{bansil13}. In many instances, the  surrounding fluids might not be totally flowing but instead possess their own relaxation dynamics, which might occur on time scales comparable to the intrinsic time scales of the   locomotion. For example, during  biofilm formation, bacteria produce an  extracellular polymeric matrix for which they control the mechanical properties through  water content  \citep{costerton95,Wilking2011} (Fig.~\ref{fig9}a).  A number of studies have attempted to quantify the impact of such complex fluids on the  swimming kinematics and energetics of bacteria.

Early studies were concerned  with perhaps the simplest question possible:  How is  bacterial locomotion modified by a change in the viscosity of the surrounding (Newtonian) fluid? Flagellated bacteria display a systematic two-phase response, where an increase in viscosity is seen to first lead to a small increase in the swimming speed followed by a sharp decrease at large viscosities  \citep{shoesmith60,schneider74,greenberg77}. The decrease has been rationalized  by assuming that the rotary  motor is working at constant power output \citep{keller74}, an assumption we now know is  erroneous since it is the motor torque which is constant  (see \S\ref{sec:overview}).     In stark contrast with flagellated bacteria, spirochetes show a systematic enhancement of their swimming speed with an increase of the viscosity of the fluid \citep{kaiser75}. In a related study,
the viscosity required  to immobilize bacteria was measured to be larger, by orders of magnitude, for spirochetes compared to flagellated bacteria \citep{greenberg77}. 

How can the difference between flagellated  bacteria and spirochetes be rationalized? One argument brought forward focused on the details of the microstructure of the fluid, arguing that all increases in viscosity are not created equal \citep{BergTurner1979}. Increases in viscosity are induced by  dissolving polymers, but the nature of the  polymer chains structure  may be critical. Experiments showing large increases in swimming speeds used solution of long unbranched chain polymers (e.g.,~methylcellulose) which are expected to form a network against which the  cells are  effectively able to push. In contrast, solutions of branched polymers are more homogeneous, and should lead to a decrease of the swimming speed since, for a constant-torque motor rotating in a viscous fluid, the product of the viscosity by the  rotational frequency needs to remain constant. The idea that  the presence of the microstructure itself in the solvent  is critical in explaining the change in swimming speeds was further modeled mathematically, with a similar physical picture  \citep{magariyama02,leshansky09}. Recent experiments further shed light on the interplay between the viscosity of the fluid and the  torque applied by the rotary motor \citep{martinez14}.  If the fluid is anisotropic (for example a liquid crystal), a strong coupling can develop between the director field and the bacteria, enabling constrained locomotion and fascinating novel flow features with truly  long-ranged interactions   (Fig.~\ref{fig9}b) \citep{zhou14}.

Recent work has focused on viscoelastic fluids with the primary motivation to understand  the locomotion of bacteria, and higher organisms,  in mucus  \citep{lauga14}. Such fluids have the added complexity of having finite relaxation times, $\lambda$, and for flagella rotating with frequency $\omega$ a dimensionless Deborah number, $\De = \lambda \omega$, needs to be considered. Small-amplitude theory on flagellar propulsion predicted that the   relaxation in the fluid would always leads to slower swimming  \citep{lauga07,fu07}. Experiments with macroscopic helices driven in rotation and undergoing force-free translation confirmed the small-amplitude theoretical results but showed that helices with larger amplitudes could experience a moderate increase in translational speed for order one Deborah numbers  \citep{Liu2011}. Detailed numerics 
on the same setup argued that the increase was due to the rotating flagellum revisiting its viscoelastic wake, as 
illustrated in Fig.~\ref{fig9}c-d for $\De =0.3$ (top) and $\De =1.5$ (bottom)  \citep{spagnolie13}.

\section{Summary Points}
\begin{enumerate}

\item Low-Reynolds-number hydrodynamics is at the heart of the fundamental physics of bacteria swimming, first and foremost by the generation of  propulsive forces. It might have also played a role in the evolution of the bacterial flagellum  resulting in optimal polymorphic shapes used for forward motion.

\item At the level of a single-cell, fluid dynamic forces are exploited by peritrichous  bacteria to induce fast bundling and unbundling of their multiple flagella, resulting in efficient reorientation mechanisms.

\item The flows created by swimming bacteria affect the interactions with their environment and with other cells leading to   enhanced nutrient transport and collective locomotion. Similarly, complex microstructures  or  external flows  can have significant influence on the trajectories, concentrations, and  orientations of swimming bacteria.

\item  Although  surrounding fluids play  important roles in the life of bacteria, hydrodynamic forces on  flagellar filaments  and cell bodies   often  have to be balanced with bending and twisting elasticity, short-range electrostatic  interactions, and are subject to both biochemical and thermal noise. Fluid dynamics is therefore only one of the    complex physical processes at play.

\end{enumerate}

\section{Future Issues}

\begin{enumerate}

\item    Why is the value of the experimentally-measured  torque applied by the rotary motor so different from that  from  hydrodynamic predictions \citep{darnton07_torque}?  What are the other mechanical forces at play?

\item    Why does the flagellar hook of peritrichous bacteria not buckle, as it does for polar bacteria under seemingly similar propulsive conditions \citep{son2013bacteria}?  Why is the flagellar hook so precisely tuned in size and rigidity that flagellar bundling  is no longer possible if small changes are made to it   \citep{brown12}?

\item   How exactly  can a tight bundle of multiple flagella   open up and close  repeatedly without jamming under the uncoordinated action of a few motors \citep{macnab1977}? How much of this process is truly driven by hydrodynamic interactions vs.~other passive mechanisms?

\item  Can we  derive predictive whole-cell models able to reproduce the full run-and-tumble statistics of swimming bacteria based solely  on our understanding on the behavior of rotary motors  \citep{turner05}?

\end{enumerate}

\section*{Acknowledgements}

I thank Tom Montenegro-Johnson for his help with the figures. Discussions with 
Howard Berg and Ray Goldstein are gratefully acknowledged. This work was funded in part by  the European Union (through a Marie-Curie CIG Grant) and by the Isaac Newton Trust (Cambridge).

\bibliographystyle{FluidMechanics}
\bibliography{annurev_refs}
\end{document}